# Images of a fourth planet orbiting HR 8799

**Christian Marois[1], B. Zuckerman[2] , Quinn M. Konopacky[3], Bruce Macintosh[3], & Travis Barman[4]**

[1]National Research Council Canada, Herzberg Institute of Astrophysics, 5071 West Saanich Rd, Victoria, BC, V9E 2E7, Canada. [2]Physics & Astronomy Department, University of California, Los Angeles, CA 90095, USA. [3]Lawrence Livermore National Laboratory, 7000 East Ave, Livermore, CA 94550, USA. [4]Lowell Observatory, 1400 West Mars Hill Road, Flagstaff, AZ 86001, USA.

**High-contrast near-infrared imaging of the nearby star HR 8799 has shown three giant planets[1]. Such images were possible due to the wide orbits (> 25 AU) and youth (< 100 Myr) of the imaged planets, which are still hot and bright as they radiate away gravitational energy acquired during their formation. A major area of contention in the extrasolar planet community is whether outer planets (> 10 AU) more massive than Jupiter form via one-step gravitational instabilities[2] or, rather, via a two-step process involving accretion of a core followed by accumulation of a massive outer envelope composed primarily of hydrogen and helium[3]. Here we report the presence of a fourth planet, interior to and about the same mass as the other three. The system, with this additional planet, represents a challenge for current planet formation models as none of them can explain the *in situ* formation of all four planets. With its four young giant planets and known cold/warm debris belts[4], the HR 8799 planetary system is a unique laboratory to study the formation and evolution of giant planets at wide > 10 AU separations.**

New near-infrared observations of HR8799, optimized for detecting close-in planets, were made at the Keck II telescope in 2009 and 2010. (See Table 1 for a summary.) A subset of the images is presented in Figure 1. A fourth planet, designated HR 8799e, is detected at six different epochs at an averaged projected separation of 0.368 ± 0.009" (14.5 ± 0.5 AU). Planet e is bound to the star and is orbiting counter-clockwise (see Figure 2), as are the three other known planets in the system. The measured orbital motion, 46 ± 10 mas/year, is consistent with a roughly circular 14.5 AU radius orbit with a ~50 year period.

Knowledge of the age and luminosity of the planets is critical for deriving their fundamental properties including mass. In 2008 we used various techniques to estimate an age of 60 Myr with a plausible range between 30 and 160 Myr, consistent with an earlier estimate of 20 to 150 Myr[12]. Two recent analyses (R. Doyon, et al. and B. Zuckerman et al. in prep) independently deduce that HR 8799 is very likely to be a member of the 30 Myr Columba Association[13]. This conclusion is based on common Galactic space motions and age indicators for stars located between the previously-known Columba members and HR8799. The younger age suggests smaller planet masses, but to be conservative, we use both age ranges (30[20-50] Myr (Columba association) and 60[30-160] Myr[1]) to derive the physical properties of planet e.

HR 8799e is located very near planets c and d in a $K_s$ versus $K_s$-L' color-magnitude diagram, suggesting that all three planets have similar spectral shapes and bolometric luminosities. We, therefore, adopt the same luminosity for these three planets; however, given the larger photometric error-bars and sparse wavelength coverage associated with planet e, we have conservatively assigned to it a luminosity error (Table 1) twice as large as those for planets b, c, and d[1]. This luminosity estimate is consistent with empirically calibrated bolometric corrections for brown dwarfs[14], although such corrections may be ill-suited for young planets with distinct spectra and colors. Using the two overlapping age ranges outlined above and the evolutionary models described in the HR 8799bcd discovery article[1], we estimate the mass of planet e to be 7[5-10] $M_{Jup}$ (30 Myr) and 10[7-13] $M_{Jup}$ (60 Myr); see Figure 3. The broadband photometry of planets b, c, and d provide strong evidence for significant atmospheric cloud coverage while recent spectroscopy of planets b and c show evidence for non-equilibrium $CO/CH_4$ chemistry[20-22]. Given the limited wavelength coverage of the discovery images for planet e, it is too early to say much about the atmospheric properties of this particular planet; however, given the similar near-IR colors as the other three planets, we can anticipate similar cloud structure and chemistry for planet e.

Stability analyses[8-10] have shown that the original 3-planet system may be in a mean motion period resonance with an upper limit on planetary masses of ~20 $M_{Jup}$ assuming an age of up to 100 Myr. With the discovery of a fourth planet, we revisit the stability of this system. We searched for stable orbital configurations with the HYBRID/Mercury package[23] using the 30 Myr (5, 7, 7 and 7 $M_{Jup}$ for b, c, d and e) and 60 Myr (7, 10, 10 and 10 $M_{Jup}$) masses. In our preliminary search we held the parameters for b, c, and d fixed to those matching either the single resonance (1:2 resonance between planets c and d only) or double resonance (1:2:4 resonance between planets b, c, and d) stable solutions found to date[10], but allowed the parameters for e to vary within the regime allowed by our observations. In 100,000 trials, seven solutions for e were found using the 30 Myr masses, based on the single-resonance configuration, that are stable for at least 160 Myr (the maximum estimated age of the system), and an additional five solutions were found that are stable for over 100 Myr. All maximally stable solutions have a semimajor axis of ~14.5 AU, with planets c, d and e in a 1:2:4 resonance (planet b not in resonance). A set of 100,000 trials were also performed using the 60 Myr masses, but only two solutions were found that are stable for over 100 Myr, each of which requires a semimajor axis of ~12.5 AU, 4-sigma away from our astrometry. This is suggestive that a younger system age and lower planet masses are preferred, although a much more thorough search of parameter space is required (see the Supplementary Information for tables of stable solutions).

The mechanism for forming this system is unclear. It is challenging for gravitational-instability fragmentation to occur at $a < 20$-$40$ AU[24,25] – ruling that out for *in situ* formation of planet e. In addition, disk instability mechanisms preferentially form objects more massive than these planets[25,26]. If the HR 8799 system represented low-mass examples of such a population, brown-dwarf companions to young massive stars would be plentiful. Nearby young star surveys[27-29] and our nearly-complete survey of 80 stars with similar masses and ages to HR 8799, have discovered no such population of brown dwarf companions. HR 8799e and possibly d are close enough to the primary star to have formed by bottom-up accretion *in situ*[24], but planets b and c are located where the collisional time-scale is conventionally thought to be too low for core accretion to form giant planets before the system's gas is depleted. A hybrid process with different planets forming through different mechanisms cannot be ruled out but seems unlikely with the similar masses and dynamical properties of the four planets. It is possible that one mechanism dominated the other and the planets later migrated to their current positions. The HR 8799 debris disk is especially massive for a star of its age or any older age main sequence star[17], which could indicate an extremely dense protoplanetary disk. Such a disk could have induced significant migration, moving planets formed by disk-instability inward or the disk could have damped the residual eccentricity from multi-planet gravitational interactions that moved core-accretion planets outward. The massive debris disk and the lack of higher-mass analogs to this system do suggest that HR 8799 represents the high-mass end of planet formation.

The HR 8799 system does show interesting similarities with the Solar system with all giant planets located past the system's estimated snow line (~2.7 AU for the Solar system and ~6 AU for HR 8799) and their debris belts located at similar equilibrium temperatures (see Figure 4). With its more massive planets, massive debris belts and large scale, the HR 8799 planetary system is also an amazing example of Nature's ability to produce extreme systems.

**Supplementary Information** is linked to the online version of the paper at www.nature.com/nature.

Acknowledgement: We thank the Keck staff, particularly H. Lewis, B. Goodrich and J. Lyke for support with the follow-up observations. We thank Gregory Laughlin & Daniel C. Fabrycky for insightful discussions. Portions of this research were performed under the auspices of the US Department of Energy by LLNL and also supported in part by the NSF Center for Adaptive Optics. We acknowledge support by NASA grants to UCLA, LLNL, and Lowell Observatory. The data were obtained at the W.M. Keck Observatory. This publication makes use of data products from the Two Micron All Sky Survey and the SIMBAD database.

**Author Contributions** The authors contributed equally to this work.

**Author Information** Reprints and permissions information is available at

www.nature.com/reprints. The authors declare no competing financial interests. Readers are welcome to comment on the online version of this article at www.nature.com/nature.
Correspondence and requests for materials should be addressed to C. M. (christian.marois@nrc-cnrc.gc.ca)

**Table 1 | HR 8799e astrometry, photometry and physical characteristics**

| Epoch | Separation w.r.t the host star in [E, N]" |
|---|---|
| 2009 July 31 Kp-band 2.124 μm (± 0.019") | [-0.299, -0.217] |
| 2009 Aug. 1 L'-band 3.776 μm (± 0.013") | [-0.303, -0.209] |
| 2009 Nov. 1 L'-band 3.776 μm (± 0.010") | [-0.304, -0.196] |
| 2010 July 13 Ks-band 2.146 μm (± 0.008") | [-0.324, -0.174] |
| 2010 July 21 L'-band 3.776 μm (± 0.011") | [-0.324, -0.175] |
| 2010 Oct. 30 L'-band 3.776 μm (± 0.010") | [-0.312, -0.151] |
| Proj. sep. (AU) – avg. from all epochs[a] | 14.5 ± 0.5 |
| Orbital Motion ("/year) | 0.046 ± 0.010 |
| Period for a face-on circular orbit (years) | ~50 |
| Planet-to-star flux ratios (diff. of magnitude)[b] | |
| ΔKs 2.146 microns (mag) | 10.67 ± 0.22 |
| ΔL' 3.776 microns (mag) | 9.37 ± 0.12 |
| Absolute magnitudes | |
| $M_{Ks}$ 2.146 microns (mag) | 12.93 ± 0.22 |
| $M_{L'}$ 3.776 microns (mag) | 11.61 ± 0.12 |
| Luminosity (log $L_{Sun}$) | -4.7 ± 0.2 |
| Mass ($M_{Jup}$) for 30[20-50] Myr | 7[5-10] |
| Mass ($M_{Jup}$) for 60[30-160] Myr | 10[7-13] |

[a] The projected separation error (in AU) also accounts for the star distance uncertainty.

[b] No reliable photometry was derived for the Kp-band 2009 July 31 data.

Figure captions

**Figure 1 | HR 8799e discovery images**. Images of HR8799 (a star at 39.4 ± 1.0 pc and located in the Pegasus constellation) were acquired at the Keck II telescope with the Angular Differential Imaging technique (ADI)[5] to allow a stable quasi-static point spread function (PSF) while leaving the field-of-view to rotate with time while tracking the star in the sky. The ADI/LOCI[5,6] SOSIE software[7] was used to subtract the stellar flux, and to combine & flux-calibrate the images. Our SOSIE software[7] iteratively fits the planet PSF to derive relative astrometry and photometry (the star position and its photometry were obtained from unsaturated data or from its PSF core that was detectable through a flux-calibrated focal plane mask). The upper left panel shows an L'-band image acquired on July 21, 2010, while the upper right panel shows a Ks-band image acquired on July 13, 2010 (arrows point toward planet e) and the bottom panel shows an L'-band image acquired on November 1, 2009. All three sequences were ~1h long. No coronagraphic focal plane mask was used on July 13 or July 21, 2010, but a 400 mas diameter mask was used on November 1, 2009. HR 8799e is located SW of the star. Planets b, c and d are seen at projected separations of 68, 38 and 24 AU from the central star, consistent with roughly circular orbits at inclinations of < 40 degrees[8-10]. Their masses (7, 10 and 10 $M_{Jup}$ for b, c, and d for 60 Myr[1], while they are 5, 7 and 7 $M_{Jup}$ for 30 Myr) were estimated

from their luminosities using age-dependent evolutionary models[11]. North is up and East is left.

**Figure 2 | HR 8799e 2009/2010 astrometry.** The 2009-2010 orbital motions of the four planets are shown in the larger plot. A square symbol denotes the first 2009 epoch. The upper-right small panel shows a zoomed version of e's astrometry including the expected motion (curved line) if it is an unrelated background object. Planet e is confirmed as bound to HR 8799 and it is moving 46 ± 10 mas/year counter-clockwise. The orbits of the solar system's giant planets (Jupiter, Saturn, Uranus and Neptune) are drawn to scale (light gray circles). With a period of ~50 years, the orbit of HR 8799e will be rapidly constrained by future observations; at our current measurement accuracy it will be possible to measure orbital curvature after only 2 years.

**Figure 3 | The mass of HR 8799e from the age-luminosity relationship.** Solid lines are luminosity-versus-age tracks for planet evolution models[11]. Objects above 13 Jupiter-mass are typically considered to be outside the planet-mass regime; however, the tail end of the planet distribution found by radial velocity surveys extends above this IAU-defined mass limit[15]. The cross hatch areas (30[20-50] Myr) and the grey boxes (60[30-160] Myr) are the adopted luminosity (1 s.d. error bars) and estimated age ranges for the four HR 8799 planets; planets c, d & e have similar luminosities, but the luminosity uncertainty for e is larger and indicated by the darker box/opposite hatch. For comparison, the ages and luminosities of recently imaged planet-mass companions near other stars are indicated (showing 1 s.d. error bars for the luminosity and estimated age ranges). An

astroseismology study suggested that the HR 8799 system might be as old as ~1 Ga[16], but it is highly unlikely such an old star would have very massive debris belts[17,18]; such an age would also require planetary masses far too high for long-term stability[10]. The older age also requires an inclination of the stellar pole relative to the line of sight of ~50 degrees, inconsistent with the nearly face-on planetary system and the ~25 degree inclination upper limit measured from Spitzer images of the outer dust halo[4]. Mass estimates based on any existing evolutionary model at ages as young as 20-30 Myr suffer from unconstrained initial formation conditions; the masses presented here could be underestimated if the planets formed by core-accretion, though "cold start" core-accretion models[19] do not reproduce the observed luminosity for any combination of mass and age. While this additional uncertainty can lead temporarily to ambiguity about the planets' masses and formation history (core-accretion or gravitation instability), it does highlight the importance of discovering and following in orbit planet-mass companions at ages when formation processes are important.

**Figure 4 | HR 8799/Solar system comparison.** The solar system (upper panel) compared to the HR 8799 system (bottom panel). HR 8799 infrared data indicate the existence of an asteroid belt analog located at 6-15 AU (we have moved the estimated outer edge of this belt to 10 AU due to e's estimated chaotic region[30]), a Kuiper-Belt-like debris disk at > 90 AU and a small particle halo extending up to 1,000 AU[4]. The red shaded regions represent the locations of the inner and outer debris belts in both systems (the Solar system Oort comet cloud and the HR 8799 halo are not shown in the above figure). The horizontal-axis of the HR 8799 plot is compressed by the square root of the

ratio of the luminosity of HR 8799 ($4.92 \pm 0.41\ L_{Sun}$) to that of the Sun to show both systems over the same equilibrium temperature range. Given the current apparent separations of the four planets of HR 8799 and the preferred locations of the inner warm debris disk and the inner edge of the outer cold disk (90 AU)[4], then (1) the 4:1 & 2:1 period resonances between the inner/outer edge of the warm debris belt and the planet e and (2) a 3:2 mean motion resonance of b with the inner edge of the outer cold disk, are both consistent with the observations. By analogy, the inner and outer edges of the main asteroid belt of our solar system are, respectively, in 4:1 and 2:1 mean motion resonances with Jupiter. Many members of the Edgeworth-Kuiper Belt, including Pluto, are in a 3:2 mean motion orbital resonance with Neptune. Solar system planet images are from NASA; HR 8799 artwork from Gemini Observatory and Lynette Cook. Planet diameters are not to scale.

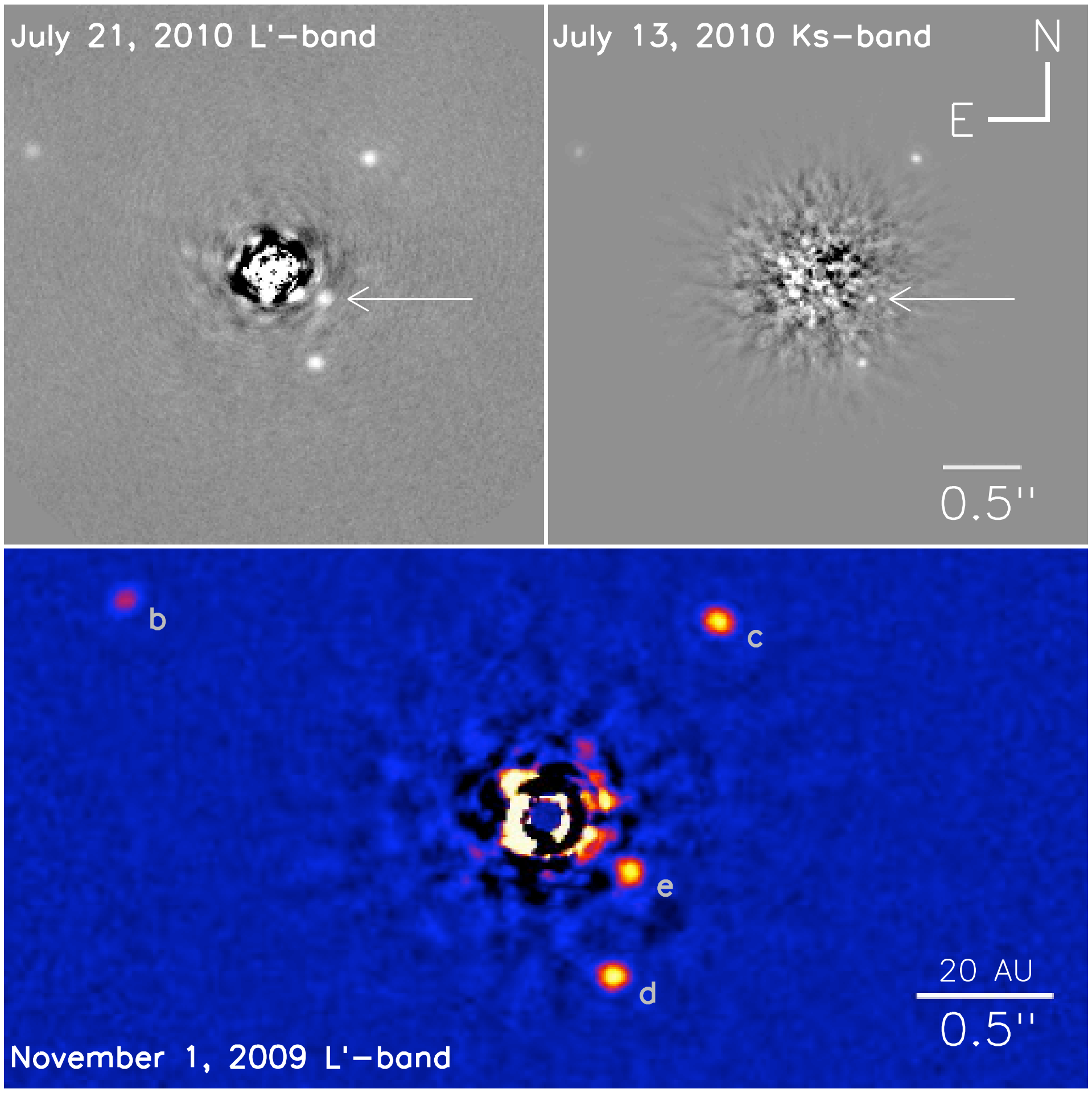
July 21, 2010 L'−band
July 13, 2010 Ks−band
N
E
0.5''
b
c
e
d
20 AU
0.5''
November 1, 2009 L'−band

**Figure 1**

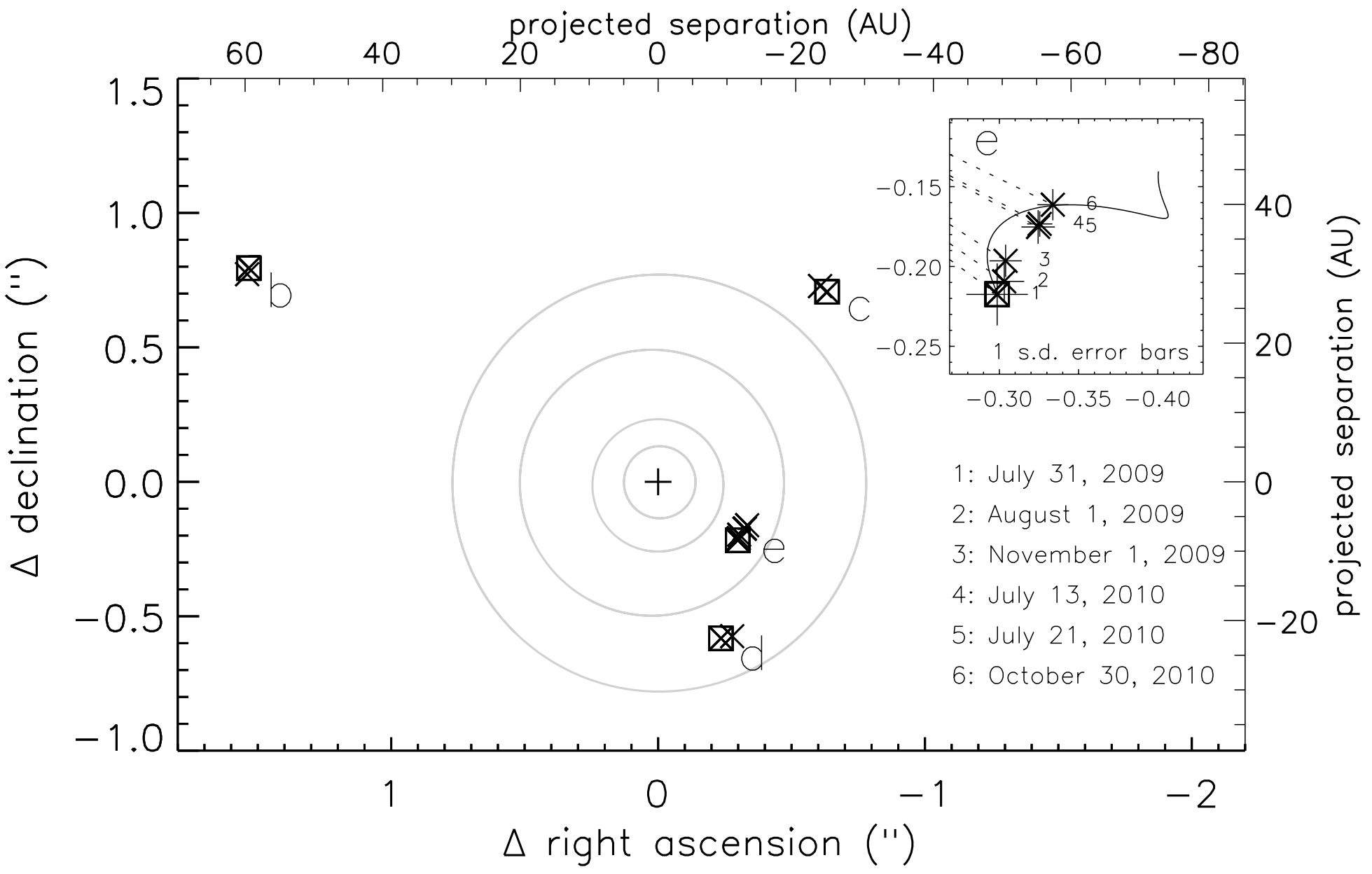

**Figure 2**

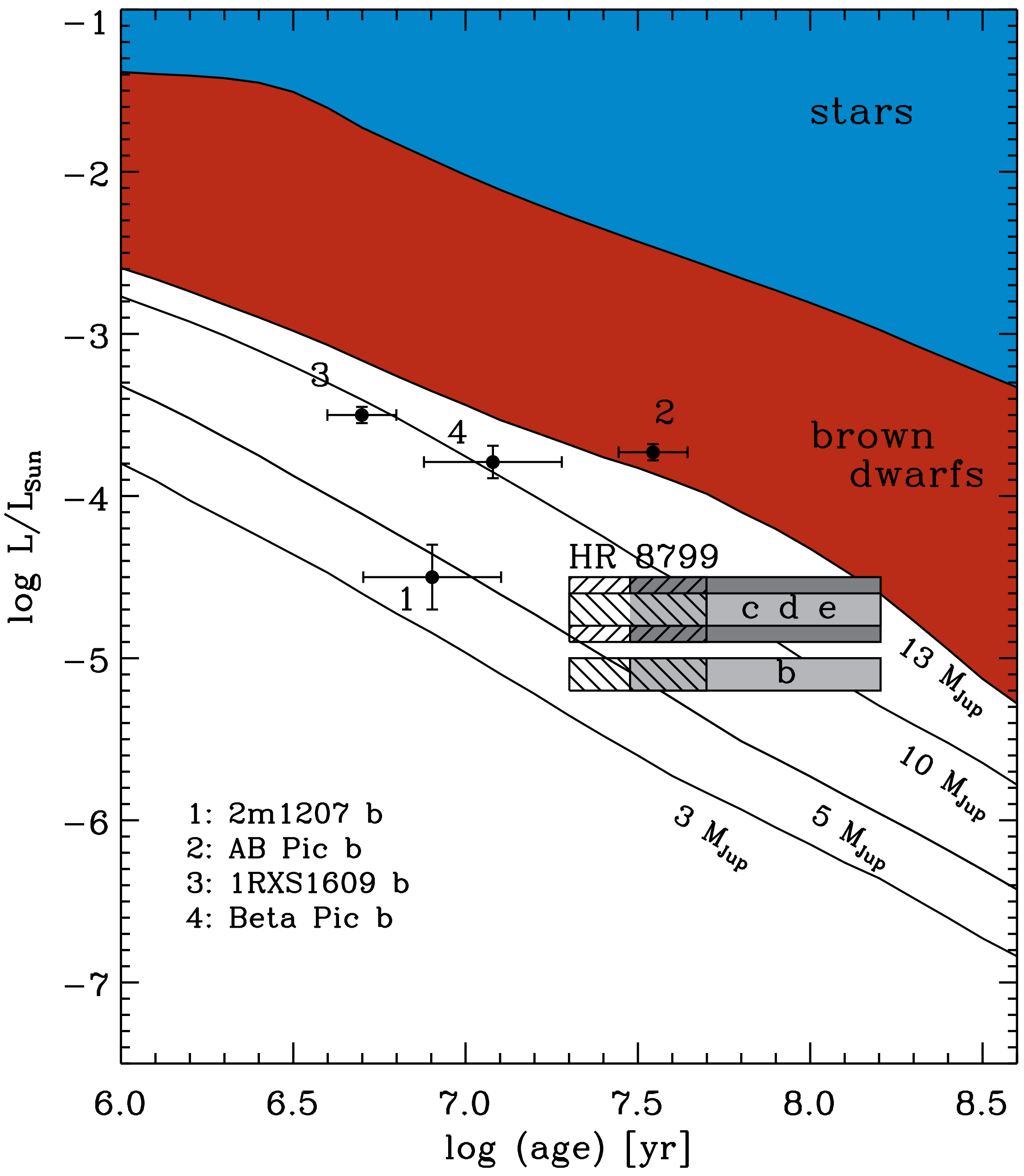

**Figure 3**

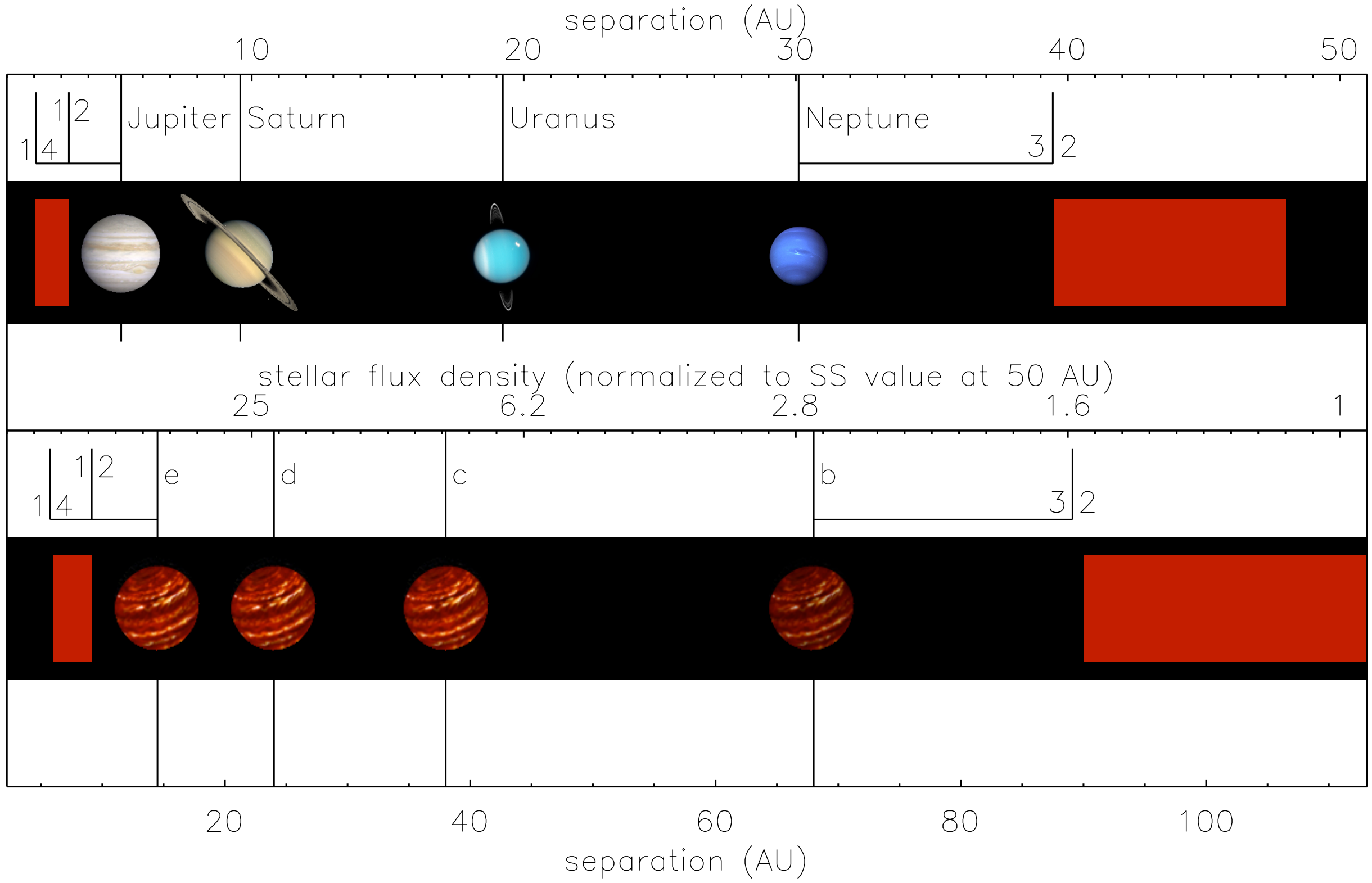

separation (AU)
10
20
30
40
50
1
4
1
2
Jupiter
Saturn
Uranus
Neptune
3
2
stellar flux density (normalized to SS value at 50 AU)
25
6.2
2.8
1.6
1
1
4
1
2
e
d
c
b
3
2
20
40
60
80
100
separation (AU)

**Figure 4**

1. Supplementary Tables

**Supplementary Table 1 | Orbital elements for planet e that are stable for at least 100 Myr for masses 5, 7, 7, and 7 $M_{Jup}$ (30 Myr age).**

| a (AU) | e | i (deg) | ω (deg) | Ω (deg) | Mean Anomaly (deg) | t stable (Myr) |
|---|---|---|---|---|---|---|
| 14.528 | 0.0676 | 0.0 | 250.900 | 0.0 | 83.305 | >160 |
| 14.501 | 0.0207 | 0.0 | 0.0 | 0.0 | 345.502 | >160 |
| 14.496 | 0.0162 | 0.0 | 0.0 | 0.0 | 345.423 | >160 |
| 14.490 | 0.0014 | 0.0 | 0.0 | 0.0 | 345.374 | >160 |
| 14.458 | 0.0501 | 0.0 | 230.036 | 0.0 | 100.012 | >160 |
| 14.434 | 0.0714 | 0.0 | 202.306 | 0.0 | 164.690 | >160 |
| 14.359 | 0.0261 | 0.0 | 243.181 | 0.0 | 74.717 | >160 |
| 14.472 | 0.005 | 0.0 | 233.930 | 0.0 | 114.566 | 143 |
| 14.499 | 0.0066 | 0.0 | 0.0 | 0.0 | 345.415 | 116 |
| 14.497 | 0.0075 | 0.0 | 0.0 | 0.0 | 345.454 | 114 |
| 14.501 | 0.0052 | 0.0 | 0.0 | 0.0 | 345.556 | 108 |
| 14.563 | 0.0483 | 0.0 | 183.213 | 0.0 | 175.416 | 103 |

**Supplementary Table 2 | Orbital elements for planet e that are stable for at least 100 Myr for masses 7, 10, 10, and 10 $M_{Jup}$ (60 Myr age)**

| a (AU) | e | i (deg) | ω (deg) | Ω (deg) | Mean Anomaly (deg) | t stable (Myr) |
|---|---|---|---|---|---|---|
| 12.437 | 0.0348 | 0.0 | 139.827 | 0.0 | 225.934 | 119 |
| 12.519 | 0.0288 | 0.0 | 32.139 | 0.0 | 339.565 | 111 |

We have followed the convention of previous authors[10] and used an arbitrary starting date, but have verified that there are epochs at which each solution is consistent with the astrometric data. Both i and Ω were fixed to 0 degrees to be coplanar with the previously-determined stable solutions for b, c, and d[10]. The dynamical simulations were stopped at 160 Myr. All orbital elements for b, c, and d are the fourth set listed in Table 3 of the single resonance case[10]. Note that while all solutions found for the lower masses are consistent with our astrometry, those found for the higher masses require planet e to be ~4σ closer to the star than our measurements.